\documentclass{article}
\usepackage{spconf,amsmath,graphicx}
\usepackage{subcaption}

\title{Is style all you need? Dependencies between emotion and GST-based speaker recognition}
\name{Morgan Sandler, Arun Ross}
\address{Michigan State University\\
\ninept\texttt{sandle20@msu.edu, rossarun@cse.msu.edu}}
\begin{document}
%
\maketitle
\begin{abstract}
In this work, we study the hypothesis that speaker identity embeddings extracted from speech samples may be used for detection and classification of emotion. In particular, we show that emotions can be effectively identified by learning speaker identities by use of a 1-D Triplet Convolutional Neural Network (CNN) \& Global Style Token (GST) scheme (e.g., DeepTalk Network) and reusing the trained speaker recognition model weights to generate features in the emotion classification domain. The automatic speaker recognition (ASR) network is trained with VoxCeleb1, VoxCeleb2, and Librispeech datasets with a triplet training loss function using speaker identity labels. Using an Support Vector Machine (SVM) classifier, we map speaker identity embeddings into discrete emotion categories from the CREMA-D, IEMOCAP, and MSP-Podcast datasets. On the task of speech emotion detection, we obtain 80.8\% ACC with acted emotion samples from CREMA-D, 81.2\% ACC with semi-natural emotion samples in IEMOCAP, and 66.9\% ACC with natural emotion samples in MSP-Podcast. We also propose a novel two-stage hierarchical classifier (HC) approach which demonstrates +2\% ACC improvement on CREMA-D emotion samples. Through this work, we seek to convey the importance of holistically modeling intra-user variation within audio samples
\end{abstract}

\begin{keywords}
emotion recognition, speaker recognition, human-computer interaction, transfer learning, affective computing
\end{keywords}

\section{Introduction}
Expressing and discerning human emotions is a significant component of human interaction \cite{darwin}. It allows humans to determine appropriate responses to social and survival-related situations. Thus, by nature emotion carries vast quantities of information in a conversation. Therefore, Human Computer Interfaces (HCIs) must be capable of identifying and responding to human emotions. For example, emotive context provides more relevant solutions in healthcare applications \cite{ehealthcare, EEG} and conversational agents \cite{convo1}. Further, understanding the contribution of emotion in speaker embeddings may enable more precise generation of emotion invariant speech which could enhance speaker recognition performance. There is enough salient emotion information in speech signals that can be extracted by automated methods \cite{jenniferwilliam}. This is the foundation of Speech Emotion Recognition (SER), the method of identifying the emotive state of an individual from an input audio sample. 

In ASR work, correlations between intra-user variation caused by emotion and high Equal Error Rate (EER) have been found \cite{pappagari2020x}. Auto-encoder reconstruction error has also been proposed as a metric to quantify emotion in speech \cite{aldeneh2021you}. In general, many deep learning SER methods are hindered by insufficient training data. Data augmentation methods have been used to combat this limitation \cite{Abbaschian2021}; however, these methods often require textual content of the speech to remove intra-emotion variation caused by spoken vowel magnitude and formant shifts \cite{Abbaschian2021}. Transfer learning approaches to SER have become popular because of these limitations. We will investigate a novel state-of-the-art ASR system's (DeepTalk) dependencies on emotion state by experimenting in SER. This paper serves as a first step towards emotion-invariant ASR.

\begin{figure*}[h]
    \centering
    \includegraphics[width=\linewidth]{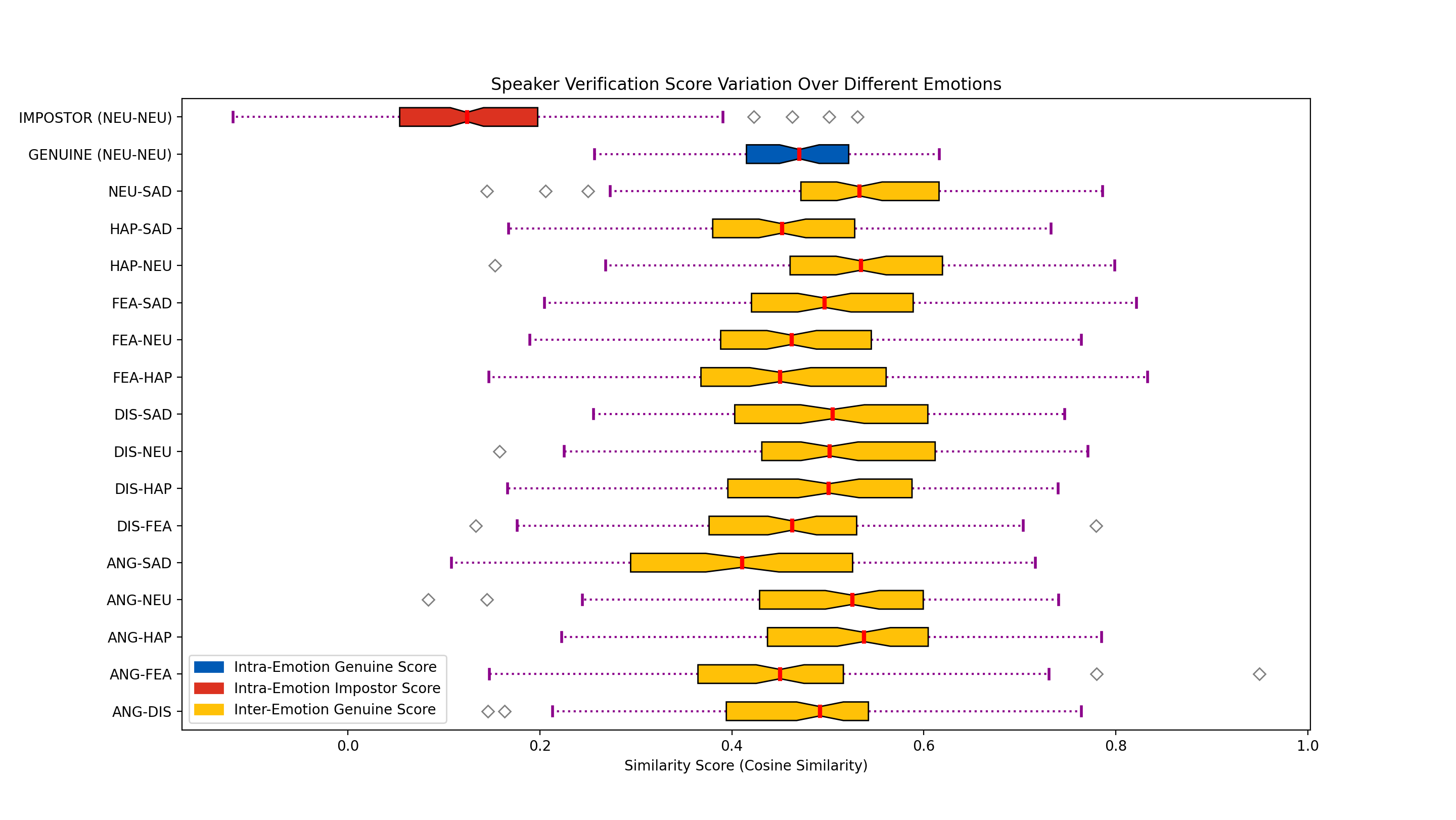}
    \caption{Box-plots demonstrating the variation of match scores caused by intra-user variation in speaker recognition resulting from change in emotion vocalization. Experiment conducted with the ECAPA-TDNN matcher \cite{ecapa} in SpeechBrain. All data is from CREMA-D \cite{crema}. We use 6 emotion classes. The same sentence is spoken but with different emotions. Speaker cosine similarity scores are illustrated across the 15 inter-emotion possibilities (of one speaker) along with the two intra-emotion tests. The baseline impostor and genuine intra-emotion box-plots are calculated by matching neutral utterances from the same and different speakers, respectively. ANG=Angry, SAD=Sad, HAP=Happy, NEU=Neutral,FEA=Fear, DIS=Disgust.}
    \label{fig:small_exp}
\end{figure*}

We hypothesize that vocal expression changes speaker-dependent attributes, consequently affecting the extracted speaker embeddings. The intuition for this hypothesis is that a speaker sounds less like themselves when emotion changes. We base this hypothesis from an experiment which demonstrates significant intra-user variation due to changes in emotion vocalization of the same textual content. We perform this experiment via another state-of-the-art speaker verification system \cite{ecapa, speechbrain}. By studying Figure \ref{fig:small_exp}, we find that there is significant intra-user variation caused by changes in emotion vocalization. The results imply that using a pre-trained model for the task of speaker verification produces speaker embeddings that may be discriminable features for use in speech emotion recognition. We analyze DeepTalk speaker embeddings in the context of emotion recognition to better understand the intra-user variation due to change in emotion state. We do so by performing an emotion classification and emotion detection experiment to make use of the variation as a predictor of emotion state.

Our work's main contributions are:
\begin{itemize}
\item{A novel use of Global Style Tokens in SER}
\item{Successful speech emotion detection and classification via transfer learning}
\item{A novel, hierarchical classifier for emotion class disambiguation}
\item{Analysis of emotion effect on speaker recognition performance}
\end{itemize}

\section{DeepTalk Encoding Network}
In this paper, we define \textit{vocal} style as a speaker's long and short term behavioral speaking habits. DeepTalk \cite{chowdhDeepTalk21} is a vocal style encoding network that captures F0 contours essential for vocal style modeling. It does so by extracting features directly from raw audio data through a 1-D Triplet CNN also referred to as DeepVOX \cite{deepvox}. DeepVOX extracts noise robust features through this method which are useful in eliminating excess intra-user variation caused by noisy audio. DeepTalk embedding features are shown to be robust to local dense regions of noise by use of dilated convolutions, effectively divulging emotion information in audio. From the DeepVOX features, a GST layer extracts DeepTalk embeddings. The GSTs identify salient style information which contain emotion content \cite{jenniferwilliam}. In this network, the DeepVOX and GST networks are trained together using a triplet-based speaker embedding learning framework to maximize the speaker-dependent vocal style information in the DeepTalk embedding. This allows DeepVOX to learn the speech representation best-suited for the speaker's vocal style characteristics through the GST network. Emotion is considered a short-term speaker attribute \cite{old_emo_st}, and would consequently be captured in the DeepTalk encoding method. These embeddings are very efficient in modeling vocal style of the speaker, only requiring one utterance of speech to stylistically reproduce a reference speaker's voice in synthesizer experiments.  \textbf{A detailed description of the DeepTalk architecture is available in \cite{chowdhDeepTalk21}}.

\section{Experiments}
\subsection{Datasets and Experimental Setup}
We use a pre-trained generic speaker DeepTalk model trained on LibriSpeech, VoxCeleb 1, and VoxCeleb 2 datasets. This model extracts fixed 256-dimensional speaker embeddings from three datasets: CREMA-D \cite{crema}, IEMOCAP \cite{iemocap}, and MSP-Podcast \cite{msp}. We have chosen an acted (CREMA-D), semi-natural (IEMOCAP), and natural (MSP-Podcast) emotion corpus to investigate how the data collection method may contribute obfuscation of emotion class. All experiment code and data partitions are detailed in our code repository\footnote{https://github.com/morganlee123/DeepTalkEmotions}.
\subsection{Feature Extraction}
The DeepTalk network is used for extracting 256-dimensional speaker embeddings for each utterance. More in-depth analysis and detail regarding the DeepTalk embedding extraction can be found in \cite{chowdhDeepTalk21}. Our implementation frame length is 22,000 and hop length is 220. This gives a 1 second frame and a hop of 10 ms. There is one emotion per audio sample. \textit{If any audio sample is less than 1 second, it is discarded}. We perform this step, to maintain stable speaker characteristics \cite{chowdhDeepTalk21}. After extracting the embeddings for each utterance, we then compute the average speaker embedding ($e_{avg}$) of $n$ utterance embeddings $e_i$.
\subsection{Classifiers}
In this work, we use SVMs with radial basis function (rbf) kernels. Our hyper-parameters are found to be C=1000 and $\gamma=0.1$ via Auto-Tuned Models \cite{ATM}. The four emotion categories for this work are: Angry, Sad, Happy, and Neutral. We choose these four emotions as they are considered the ``basic" emotions that cover the most frequent human-interactions \cite{basicemotion}. Categorical/discrete emotion classes have limitations in expressivity. For example, differentiating between specific emotions (e.g., cold anger vs hot anger). Continuous emotion dimensions are used in literature to address this problem. In spite of their limitations, we use categorical emotions for their simplicity and to allow classes between datasets to match. We design two classifiers for our experiments. A single 4-class SVM serves as our baseline experiment. Then, we employ a hierarchical classifier consisting of two SVMs in sequence: the first distinguishes the Sad emotion from the other categories, while the second distinguishes between the Angry, Happy and Neutral categories. We use this hierarchical approach to initially differentiate the most ``challenging" emotion from the rest. For example, Sad may be difficult to differentiate from the Neutral state \cite{psych_contours}. Therefore, in hierarchical classifier experiments, we first distinguish Sad, then classify the remaining emotions. These two types of speech emotion classifiers are illustrated in Figures \ref{fig:architecture} and \ref{fig:archidual}. 
\subsection{Architecture}
\begin{figure}[h!]
    \centering
    \includegraphics[width=\linewidth]{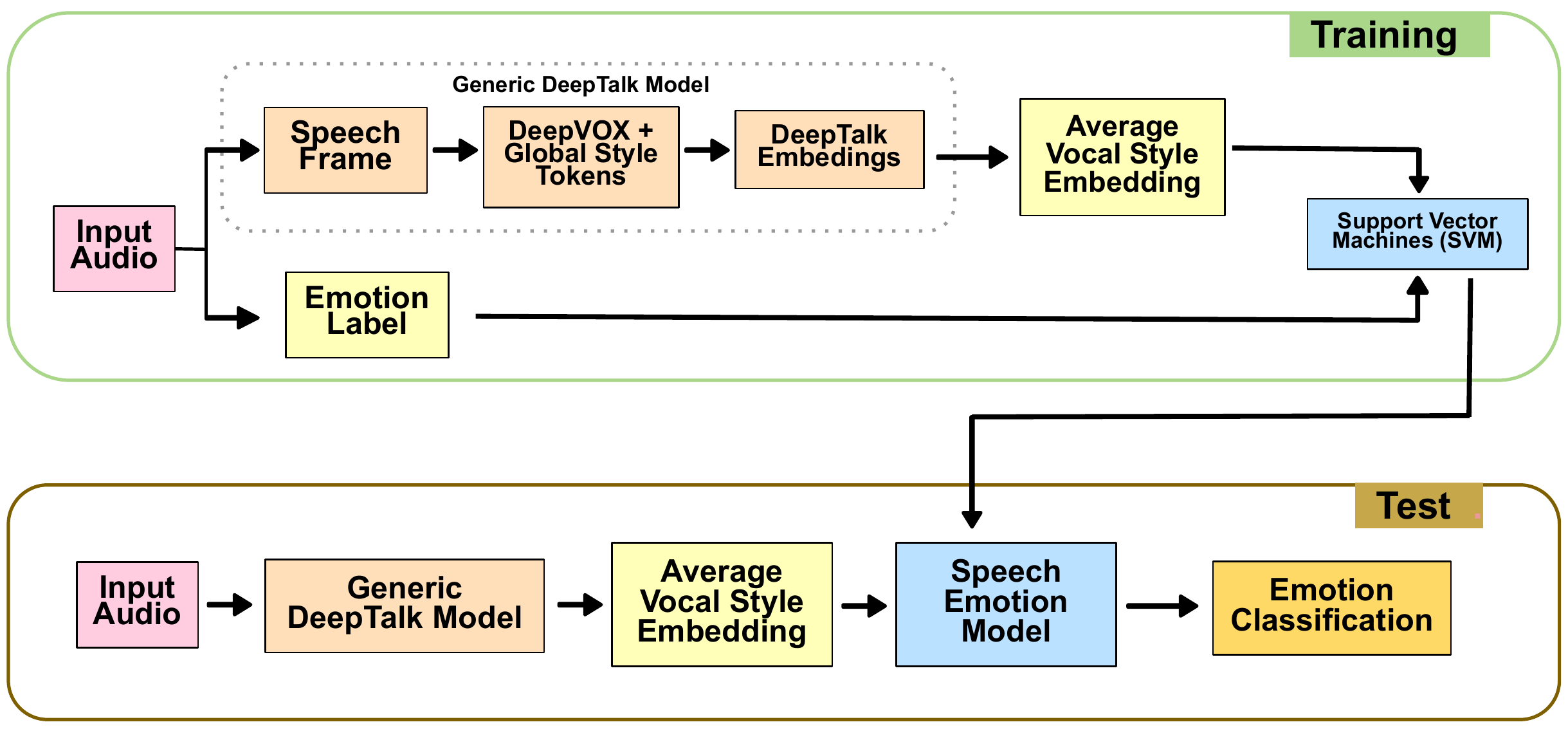}
    \caption{The overall architecture of the proposed method.}
    \label{fig:architecture}
\end{figure}

\begin{figure}[h!]
    \centering
    \includegraphics[width=\linewidth]{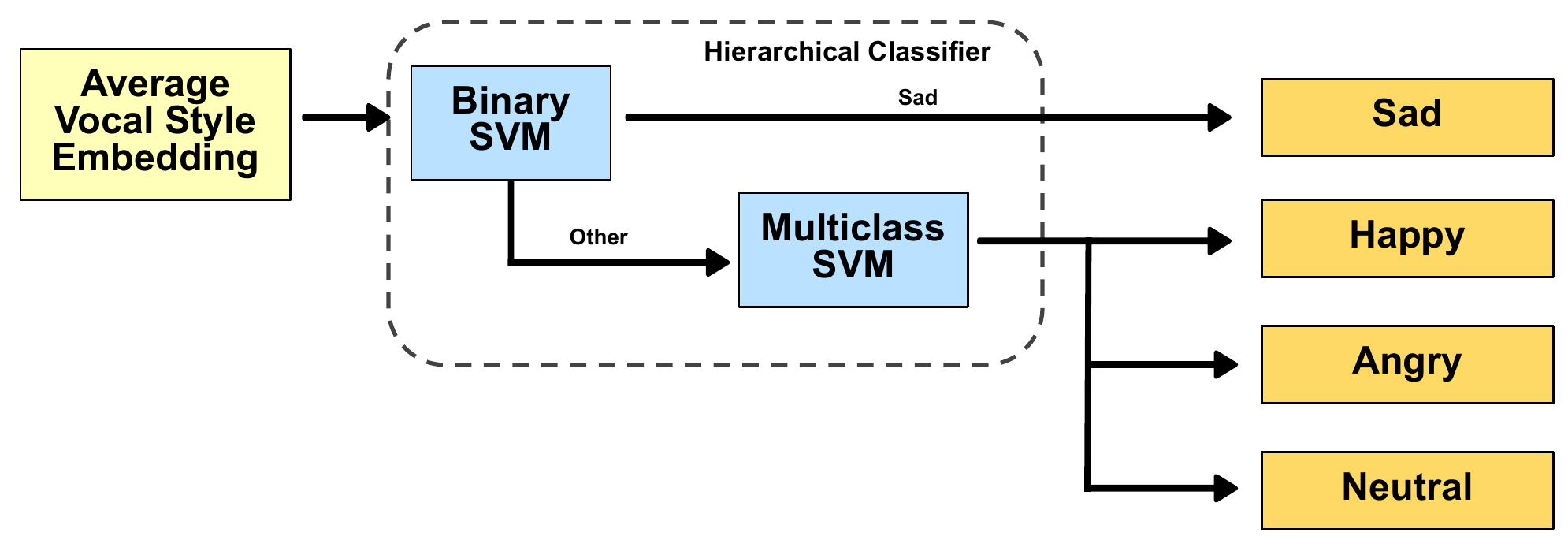}
    \caption{The hierarchical classifier (Sad-First). The first classifier is a binary classifier that discriminates a given emotion class from the rest. Those which do not belong to the first class are classified into the remaining three by the second layer.}
    \label{fig:archidual}
\end{figure}

Our general network architecture is illustrated in Figure \ref{fig:architecture}. As our baseline experiment, we use a single SVM layer. In our hierarchical classifier experiment, we propose a two-stage hierarchical SVM classifier as demonstrated in Figure \ref{fig:archidual}.

\begin{figure}[t]
    \centering
    \includegraphics[width=\linewidth]{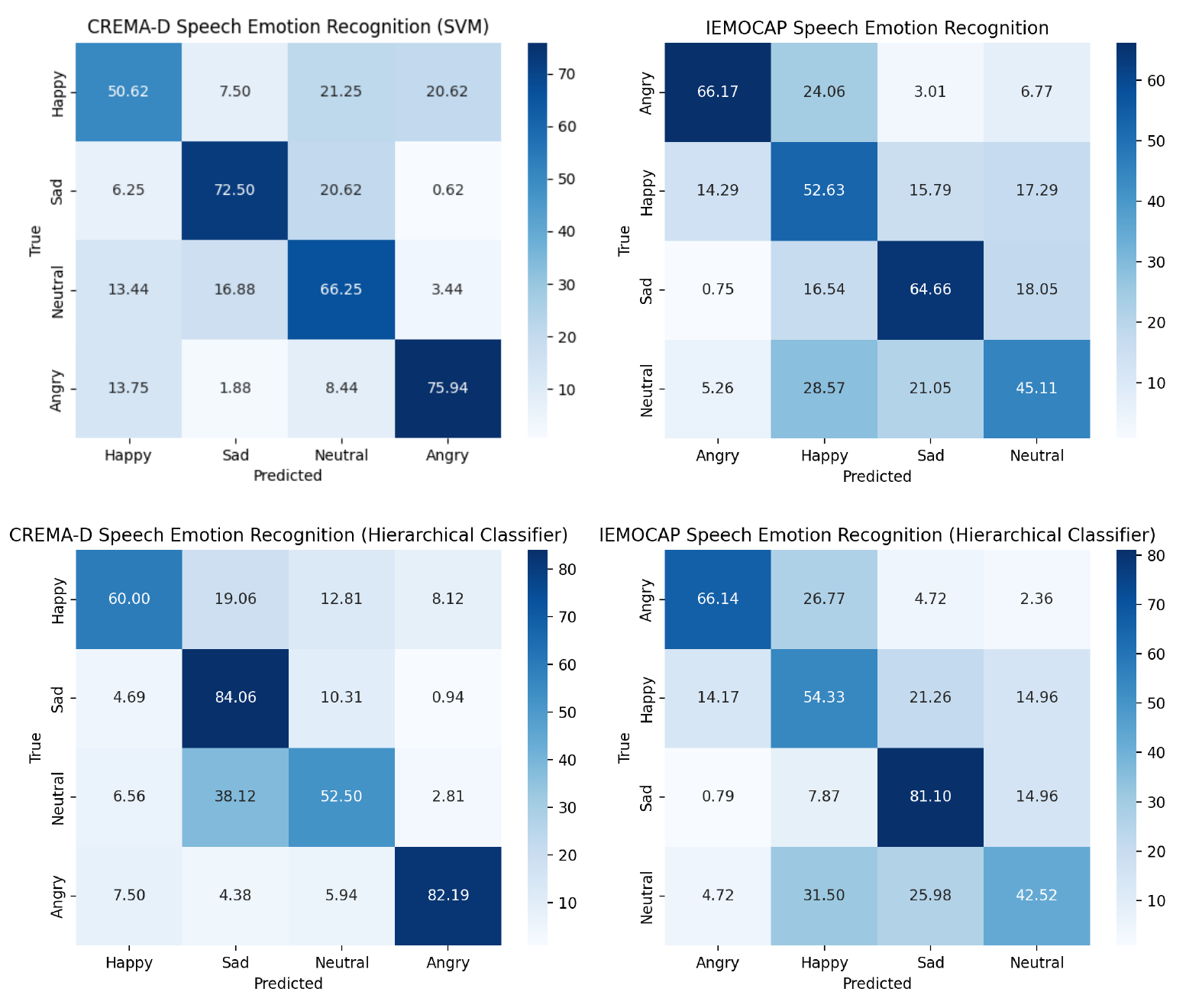}
    \caption{Experiment Results - Confusion Matrix of CREMA-D, IEMOCAP 4-class classification with an SVM and a Hierarchical Classifier}
    \label{fig:crema_conf}
\end{figure}
\begin{figure}[t]
    \centering
    \includegraphics[width=\linewidth]{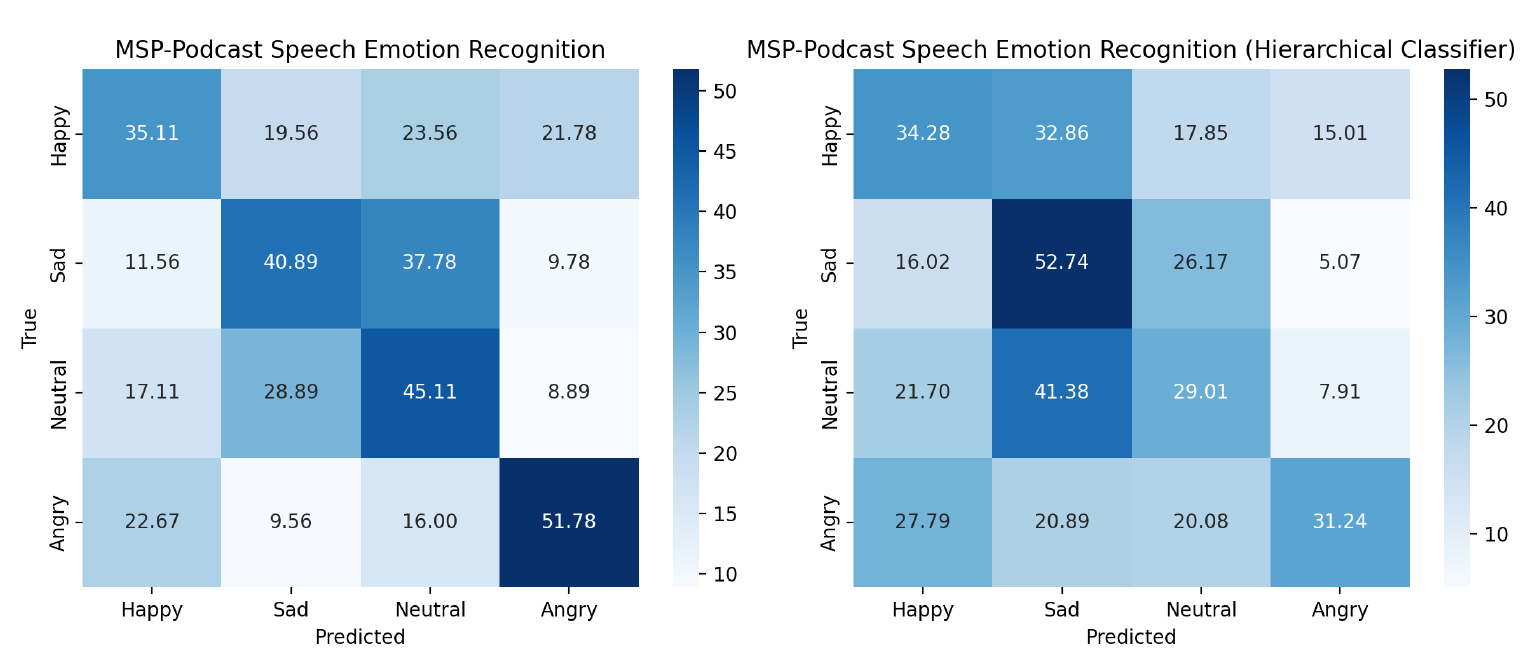}
    \caption{Experiment Results - Confusion Matrix of MSP-Podcast 4-class classification with an SVM and a Hierarchical Classifier}
    \label{fig:test}
\end{figure}

\section{Results}
From the findings in all experiments (Figures \ref{fig:crema_conf} and \ref{fig:test}), we conclude that speaker embeddings with vocal style representation can be used for the purpose of speech emotion recognition with efficacy. We do note limitations of these embeddings for this problem. Happy vocal style can be commonly confused with anger. This, could be partly explained by similarity of vocal intensity (arousal) from both emotions \cite{dimensions}. Differentiation between sad and neutral is a challenging problem, especially within natural emotion audio samples such as those from MSP-Podcast. This may relate to the cognitive appraisal used in CREMA-D to consciously produce an acted, distinct sad emotion. This subtle difference may not provide enough change in speaker characteristics to properly classify. Emotions may be significantly obfuscated by the lack of conscious vocal expression which violates our assumption that the emotion is stylistically produced through speech. 

We compare our work to a similar transfer learning approach \cite{jhu} (viz., Table \ \ref{tab:bigger_tab} \& \ref{tab:results_comparison_table}). The authors present a similar technique which they use a pre-trained ResNet \cite{jhumit} for the task of ASR. They fine-tune their model with emotion samples on each dataset. Even though this is not a perfect 1-to-1 comparison due to data used in pre-training or fine-tuning\footnotemark[2], we note that our Sad-First HC performs consistently across all emotion classes, where x-vectors tend to favor Anger and Neutral classes.\footnotetext[2]{The authors do not discard samples due to their length and the authors augment the model with generated samples to improve model robustness}

We also compare our work to another transfer learning approach, yet again pre-trained with different partitions of speaker identity data. The authors report a 4-class emotion recognition performance of 57.9\% which our Sad-First HC slightly outperforms with accuracy of 58.3\%.

\begin{table}[th]
  \caption{Performance over two tasks: 4-class emotion classification and emotion detection. Comparison of Hierarchical Classifier to baseline SVM. Metric is standard accuracy-- higher is better. ER = 4-class Emotion Recognition, ED = Emotion Detection (Neutral or Emotion-Present Classification). '-' denotes no comparison provided.}
    \label{tab:bigger_tab}
    \centering
    \resizebox{\columnwidth}{!}{%
    \begin{tabular}{||c c c c||} 
     \hline
     Algorithm & CREMA-D & MSP-Podcast & IEMOCAP  \\ [0.5ex]
     & ER / ED & ER / ED  &ER / ED \\
     \hline\hline
     Our SVM  & 66.5 / 78.9 & \textbf{43.3} / \textbf{66.9} & 57.5 / \textbf{81.2} \\ 
     Our Sad-First HC & \textbf{68.7} / \textbf{80.8} & 36.8 / 66.2 & \textbf{58.3} / 76.3 \\
     DNN ASR + SVM \footnotemark[3] \cite{aldeneh2021you} & - / - & - / - & 57.9 / - \\  [1ex] 
     \hline
    \end{tabular}
    }
 
\end{table}\footnotetext[3]{Authors perform 30 trials and select the best result}

\begin{table}[th]
  \caption{Class-wise f1-scores comparison of our Sad-First Hierarchical Classifier network to ResNet x-vector SER implementation on CREMA-D in \cite{jhu}. Note: This is not a perfect 1-to-1 comparison because we drop audio samples below a certain length to maintain integrity of speaker characteristics in our features. }
    \label{tab:results_comparison_table}
    \centering
    \resizebox{\columnwidth}{!}{%
    \begin{tabular}{||c c c c c||} 
     \hline
     Algorithm & Anger & Sad & Happy & Neutral  \\ [0.5ex]
     \hline\hline
     Our Sad-First HC &\textbf{84.7} & \textbf{68.4} & \textbf{63.7}  & $57.8$\\ 
     Pre-Trained ResNet ASR + X-Vectors\footnotemark[4] \cite{jhu} & $75.8$ & $22.4$ & $54.5$ & \textbf{88.1} \\ [1ex] 
     \hline
    \end{tabular}
    }
        
\end{table}
\footnotetext[4]{Authors perform 5-fold cross validation. There is also significant difference in data used for pre-training and fine-tuning. Details in \cite{jhu, jhumit}}

\section{Summary}
Speech Emotion Recognition from audio is a challenging problem. We show that speaker embeddings capture emotion as part of the speaker identity. We theorize that speaker identity changes with emotion state, and demonstrate an effective, simple classification model based upon that hypothesis. We also propose a simple hierarchical classifier based on these features to disambiguate between challenging emotion categories. We show that DeepTalk embeddings contain significant, inherent emotion representation. After training a GST-based model for the purpose of automatic speaker recognition, we are able to extract speaker features that are useful towards emotion classification. The hierarchical classifier method, further distinguishes similar emotion styles, leading to competitive SER performance using speaker identity features. We hope these developments motivate new lines of research in emotion modeling to improve automatic speaker recognition performance. Future directions of this work involve expanding the speaker recognition feature set to capture emotion state in speech. This may further solve the problems of False Accepts/Rejects to a Biometric authentication system. We also hope to investigate further use of continuous dimensions to classify emotion with speaker embeddings. Naturalistic datasets provide us with ``in-the-wild" emotion; therefore, we would like to perform future experiments on exploring the differences between natural emotions and acted emotions through continuous dimensions.

\bibliographystyle{IEEEtran}
\bibliography{mybib}

\end{document}